# Structural and effective charge determination for planar-zigzag β-poly(vinylidene fluoride)

*Nicholas J. Ramer\* and Kimberly A. Stiso*

*Department of Chemistry, Long Island University – C. W. Post Campus,*

*Brookville, New York 11548-1300*

(Received)

**ABSTRACT**

We present the computation of structural and polarization for a ferroelectric polymer, specifically, the β-phase of poly(vinylidene fluoride) or $(-CH_2-CF_2-)_n$. Two structures have been described in the literature: one containing an undistorted carbon backbone (planar-zigzag) and the other containing alternatively deflected $-CF_2-$ groups. Previous studies have indicated that these deflections have no significant effect on calculated elastic properties. We employ density functional theory within the generalized gradient approximation to find relaxed atomic positions for the planar-zigzag structure. The relaxed structure is in good agreement with the X-ray studies. In addition, we have computed the spontaneous polarization and Born effective charges for this structure using a Berry-phase approach. The spontaneous polarization agrees well with results from other density functional theory studies. The effective charges are considerably smaller as compared to those in inorganic ferroelectric materials and show a small amount of charge transfer between the $-CH_2-$ and $-CF_2-$ moieties.

1. INTRODUCTION

Poly(vinylidene fluoride) or PVDF has long been known to be piezoelectric.[1] PVDF exhibits a much stronger piezoelectric response than most other known ferroelectric polymers. As compared to ceramic piezoelectric materials, PVDF has a relatively low dielectric constant and low elastic stiffness. Combined these two properties result in high voltage sensitivity and low acoustic impedance for this material.

PVDF has a monomeric repeat unit of ($-CH_2-CF_2-$). There are four known crystalline forms of PVDF: $\alpha$ (Form II), $\beta$ (Form I), $\gamma$ (Form III) and $\delta$ phases (Form IV or II$_p$). The phases differ in crystallographic space group, number of formula units per unit cell, packing of polymer chains and patterning of $-CH_2-$ and $-CF_2-$ conformations. The observed conformations are as follows: *trans* or *t* (F and H atoms at 180º to polymer chain) and *gauche$^\pm$* or *g$^\pm$* (H and F atoms at ± 60º to polymer chain).[2] The angles observed experimentally are slightly different.

For example, the $\alpha$-phase shows a *tg$^+$tg$^-$* conformation pattern with four formula units per unit cell and monoclinic space group $P2_1/c$ ($C_{2h}^5$).[3,4] The chains in $\alpha$-PVDF are packed in an anti-parallel manner, yielding a non-polar structure. The $\beta$-phase possesses all-*trans* conformations with two formula units per unit cell and an orthorhombic space group and parallel chain packing (see below). The crystal structure of the $\gamma$-phase has been characterized by either an orthorhombic unit cell with space group $C2cm$ ($C_{2v}^{16}$) or monoclinic unit cell with space group $Cc$ ($C_s^4$).[5,6] Regardless of unit cell geometry, a *tttg$^+$tttg$^-$* conformation pattern has been observed with eight formula units per unit cell. The designation of the $\delta$-phase as II$_p$ represents its identification as the polar form of the non-polar $\alpha$-phase. It possesses parallel packing of polymer chains with its dipoles oriented in the same direction. Both the β- and δ-phases are ferroelectric and exhibit piezoelectricity.[2] Nine stable crystalline forms of PVDF (including the above four) have been proposed based on force-field calculations.[7]

The crystal structure of β-phase of PVDF deserves particular attention due to differing structures in the literature. Galperin *et al.* suggested planar-zigzag conformations for all carbons in the polymer (so-called all-*trans*) with an identity period of 2.55 Å.[8,9] The planar notation refers to all the carbon atoms lying in the *yz* plane. Cortili and Zerbi also indicated a planar-zigzag structure.[10] An orthorhombic unit cell with the space group *Cm2m* ($C_{2v}^{14}$) was proposed by Lando *et al.* [$a$ = 8.47 Å, $b$ = 4.90 Å, and $c$(fiber or chain axis) = 2.56 Å] containing fully planar-zigzag conformations (see Figure 1a).[11] Both the space group and lattice constants are identical to those found in previous studies.[8] They confirmed this structure by studying nuclear magnetic resonance spectra for the second moment of a β-PVDF film as a function of the angle of the magnetic field and the draw direction. They found a minimum of the second moment at 45º, indicating planar-zigzag conformations.

Subsequently, a slightly different orthorhombic unit cell was proposed by Hasegawa *et al.* [$a$ = 8.58 Å, $b$ = 4.91 Å, and $c$ = 2.56 Å] also containing fully planar-zigzag conformations.[12] In addition, they determined a structure of the unit cell with alternatively-deflected zigzag molecular conformation (see Figure 1b). The deflections cause a deformation of the polymer chain in an alternating pattern. These –$CF_2$– deflections were originally proposed by Galperin *et al.* and account for the smaller fiber axis length (2.56 Å) as compared to twice the covalent van der Waals radius of fluorine ($2 \times 1.35$ Å = 2.70 Å).[8] By including these deflections, the *c*-axis length must be doubled (see Figure 1b). The –CH$_2$– groups however remain un-deflected. However, upon examination of the X-ray fiber pattern on a highly-oriented crystalline sample, Hasegawa *et al.* did not find a layer line corresponding to a 2 × 2.56 Å = 5.12 Å period. In order to reconcile the discrepancy between the fiber-axis length and the presence of the –CF$_2$– deflections, they propose a statistically-disordered model, with the deflected chain and its mirror image occurring with equal probability in the lattice. Evidence for their statistical model was given by an improvement in the discrepancy factor, *R*, for their proposed structure as compared to the X-ray diffraction data. For their proposed planar-zigzag structure, they reported an *R* of 18.5%. By varying the –CF$_2$– deflection angle, σ, in their statistical model, a minimum *R* of 13.5% was found at σ=7º.

To further confirm the existence of the –CF$_2$– deflections, Hasegawa *et al.* computed an intermolecular potential energy for the alternatively-deflected zigzag chain as a function of σ.[13] They found a minimum potential energy for σ=7º, coinciding with the deflection angle that gave the minimum discrepancy factor in their X-ray analysis.

The inclusion of these deflections in computational studies of β-PVDF is limited. Using a point-charge model, Tashiro *et al.* studied the elastic and piezoelectric constants for alternatively-deflected β-PVDF. In order to assess the effect that the –CF$_2$– deflections have on mechanical properties of the polymer, they computed the anisotropy of the Young's modulus in the *xy* plane for β-PVDF polymers with a deflection angle of ±7º.[14] They found almost the same results for anisotropy as the planar-zigzag structure. In addition, they computed the Young's modulus in the chain direction for both planar-zigzag and deflected structures. For the planar-zigzag model, a modulus of 237 GPa was found, while for the deflected polymer, a modulus of 222 GPa was determined. It was their conclusion that these deflections have little effect on the computed mechanical properties of the polymer. In the recent study by Duan *et al*. investigated an alternatively-deflected structure for their first-principles band structure calculations using full-potential linear-augmented plane-wave method (FLAPW).[15] Due to the doubled *c*-axis length, their structure must have a modified space group. By computing the total energy for two isomorphic space groups, they determine the space group for the alternatively-deflected structure to be *Cc*2*m* instead of *Ic*2*m*. Their atomic coordinates were found by fitting bond lengths and angles to the values determined by the force-field structural optimization of Karasawa and Goddard.[7]

The ferroelectricity in PVDF results from the presence of the highly electronegative F atoms, which yield highly polar bonds with C. Coupled with the much less polar bonds between C and H, a spontaneous polarization is generated normal to the carbon-carbon bonds of the polymer chain (along the *b*- or *y*- axis) in the case of β-PVDF.

Different methods have been utilized to determine theoretically the spontaneous polarization ($P_s$) in β-PVDF. Using tabulated bond dipole moments and isotropic continuum dielectric theory, Broadhurst *et al*. first modeled the β-PVDF polarization.[16] They were unable to compute $P_s$ and relied on experimental values to calibrate their model. Subsequently, Purvis and Taylor showed that isotropic continuum dielectric theory was inaccurate in computing the β-PVDF polarization.[17] Instead, they used and orthorhombic lattice of point dipoles and received a value for $P_s$ of 0.086 C/m$^2$. Building on the work of Purvis and Taylor, Al-Jishi and Taylor incorporated not only the crystal structure but also the finite separation between monopoles to compute the $P_s$.[18] To include this effect, pairs of point charges were employed to represent the dipole moment at each lattice site and Coulomb sums to determine the local electric field. Using this method, a $P_s$ of 0.127 C/m$^2$ was found for β-PVDF.[19] These calculations were dependent on the choice of point charge separation and did not include dipole oscillations. By means of force fields derived from the vibrational spectrum of β-PVDF, Tashiro *et al*. applied molecular simulations to compute a $P_s$ of 0.140 C/m$^2$.[14] This $P_s$ does not take into account effects of electronic polarization and dipole oscillations. Also employing a force field, Karasawa and Goddard combined vibrational spectra with electronic structure calculations so as to include effects of electronic polarization but omitting dipole oscillations.[7] Carbeck *et al*. developed a model of polarization in β-PVDF from an atomic potential energy function based on the shell model of electronic polarization. From minimizing the Gibbs' free energy calculated from consistent quasi-harmonic lattice dynamics, a $P_s$ at 0 K is found to be 0.182 C/m$^2$.[20,21] Lastly, a rigid-dipole model provided a value of 0.131 C/m$^2$ for the spontaneous polarization.[22] Since typical samples of PVDF are partially crystalline and may consist of a mixture of both α- (non-polar) and β-phases (polar), experimental values for the spontaneous polarization in real samples (approximately 50% crystallinity) are smaller (0.05 to 0.08 C/m$^2$). The range of values is due to the variations in sample preparation, degree of crystallinity and experimental method for determining the polarization.[19,23,24]

Each $P_s$ can be written as a sum of electronic polarization ($P_{el}$) and ionic polarization ($P_{ion}$). $P_{ion}$ is a lattice summation according to

$$P_{ion} = \frac{e}{V} \sum_m Z_m \cdot r_m$$

where $V$ is the unit cell volume, $Z_m$ is the core charge of atom $m$ and $r_m$ is the position of the $m$-th atom in the cell. The computation of the $P_{el}$ from density-functional calculations has only recently been made possible.[25,26,27] Operationally, the $P_{el}$ or "Berry-phase" term (Ref. 28) can be computed by finite differences on a fine $k$-point grid in the Brillouin zone (BZ).[27,29] $P_{el}$ is defined as

$$P_{el} = -\frac{2ie}{8\pi^3} \sum_i \int_{BZ} dk \langle \phi_{ik} | \nabla_k | \phi_{ik} \rangle,$$

where $\phi_{ik}$ are the occupied cell-periodic Bloch states of the system. This approach is more robust that the aforementioned point-charge and dipole-dipole models. Higher order effects are automatically incorporated into this method based on the entire charge density within a unit cell. Recently, Nakhmanson *et al.* employed a multigrid-based total-energy density-functional method to compute $P_s$ using the Berry-phase approach for the $P_{el}$ term.[30] They reported a spontaneous polarization of 0.178 C/m².

Atomic charges can provide an atomistic understanding of the underlying basis for the spontaneous polarization along with various other physical phenomena, which are central to many solid-state measurements. However, the ubiquitous usage of atomic charges has also led to many definitions and approaches to their computation. These approaches unfortunately are not equivalent.[31]

The atomic charge definitions can be categorized in either static, based upon the partitioning of the charge density into contributions from specific atoms, or dynamic, defined by the change in polarization created by atomic displacement.[32]

A static atomic charge relies on the idea that the charge associated with an isolated atom is well-defined. Specifically, the delocalized electronic charge density is mapped onto localized point charges for each atom. This mapping can only be done unambiguously if and only if a "boundary can be drawn between ions so as to pass through regions which the electron density is small compared with the reciprocal of volume inclosed."[33] In the case of strong covalent bonding, as is the case in organic-based polymers, these boundaries are not clearly defined.

In computing accurate atomic charges, it is imperative that a more fundamental and intuitive method be introduced so that no ambiguity remains when assigning charge to a given atom within the system. When an atom is displaced, its charge density will also be displaced relative to the atomic displacement. An induced change in polarization will accompany the charge density displacement. This is the underlying concept of a dynamic charge, one that changes with atomic displacement.

Dynamical Born effective charges ($Z^*$) can be computed from differences in the spontaneous polarization due to small atomic displacements away from a relaxed structure. $Z^*$ for atom $m$ is defined as

$$Z_m^* = \frac{\Delta P_s}{u_m} = \frac{P_s^{\text{disp}} - P_s^{\text{undisp}}}{u_m}$$

where $u_m$ is the displacement of atom $m$, $P_s^{\text{disp}}$ is the spontaneous polarization for the displaced structure and $P_s^{\text{undisp}}$ is the spontaneous polarization for the un-displaced structure. Determination of Born effective charges computed using the Berry-phase approach has focused on inorganic perovskite materials, which contain primarily ionic bonding. The application of this approach to mostly covalently bonding materials, such as ferroelectric polymers, has not been reported.

Since there exists experimental evidence for two slightly different structures of β-PVDF, a density-functional theory determination of the atomic positions for each structure is warranted. In addition, by computing polarization properties for each structure, it will be possible to ascertain the effect the –$CF_2$– deflections have on electron distribution in the polymer. These studies can also be instrumental in proposing alternative PVDF-based polymers by atomic substitution, similar to the studies of Nakhmanson *et al*.[30] In the present study, we have chosen to complete full atomic relaxations for the planar-zigzag structure and compute its polarization and effective charges. We will employ the Berry-phase approach to the computation of the polarization and subsequently, the dynamic Born effective charges. The considerable covalent bonding within β-PVDF underscores our choice of effective charge definition and computational approach. These effective charges can be subsequently incorporated into the calculation of the β-PVDF vibrational spectrum so as to include splitting of the infrared-active phonons due to longitudinal vibrations. A companion study of the alternatively-deflected structure will be published elsewhere.[34]

## 2. METHODS

We have applied density functional theory within the generalized-gradient approximation (GGA) (Ref. 35) and optimized pseudopotential generation methods were used.[36] The parameters used in the pseudopotential construction are contained in Table I. A plane-wave cut-off energy of 50 Ry was used.[37] Since these pseudopotentials possessed a high degree of transferability, designed nonlocal pseudopotential methods were not employed.[38] To determine the accuracy of these pseudopotentials, density-functional calculations within the GGA for various small molecules were completed. From these calculations, bond lengths and vibrational frequencies were determined and found to be within the expected error range for the GGA as compared to experiment.[39,40]

For the planar-zigzag β-PVDF unit cell, the lattice constants [$a = 8.58$ Å, $b = 4.91$ Å, and $c = 2.56$ Å] and the space group $Cm2m$ ($C_{2v}^{14}$) as determined by Hasegawa *et al.* were used.[12] The unit cell contains two formula units (Z=2), one at the origin and the other at (½,½,0).

For the atomic-relaxation density-functional calculations, Brillouin zone integrations were done using a $4 \times 4 \times 4$ Monkhorst-Pack *k*-point mesh.[41] This yields 8 k-points in the irreducible wedge of the Brillouin zone. Initially, the atomic positions of planar-zigzag β-PVDF from Hasegawa *et al.* were used.[12] Atomic relaxations were completed by minimizing the computed Hellmann-Feynman forces using a conjugate-gradient algorithm on all atomic positions that are not constrained by symmetry.[42] Converged atomic positions were achieved when the forces on all atoms were less than 0.01 eV/Å. To determine the spontaneous polarization, $P_s$, an increased *k*-point mesh density must be employed parallel to the direction of polarization. For β-PVDF, converged $P_s$ and $Z^*$ were found for a $4 \times 8 \times 4$ *k*-point mesh. The Berry-phase method was employed to compute the values.[27,29]

In order to compute the spontaneous polarization using the Berry-phase method, a nonpolar reference structure must be used. This nonpolar structure is used to compute the ionic portion of the polarization. Since there are two formula units within the β-PVDF unit cell, it is possible to rotate one formula unit by 180º around the carbon chain. We have chosen instead to use a high symmetry structure that restores the mirror-plane symmetry of the *y*-axis. This structure is similar in context to the nonpolar cubic reference structures used when computing polarization in ferroelectric perovskites.

## 3. RESULTS AND DISCUSSIONS

### A. Atomic positions

The converged atomic positions are reported in Table II. C1 and C2 are the carbons attached to the H and F atoms respectively. In addition, experimentally-determined atomic positions are also included.[11,12] The relaxed atomic positions fall within the range of the experimental values for the structures.

As a further means of comparison, Table III reports bond lengths and bond angles for the theoretical and experimental structures. For the experimental structures, values are given based upon the reported atomic positions and those used to fit the X-ray patterns and intensities (in parentheses). Since H atom positions were not given in the Lando *et al.* X-ray structure (Ref. 11), C1–H bond length and H–C1–H bond angle are not given although fitting values are provided. Several considerations must be made to compare the experimental quantities to theoretical values. First, the theoretical atomic positions are considered to be at 0 K. The experimental atomic positions are found at a finite temperature above 0 K where thermal broadening can cause some variation in the assignment of the positions. In addition, the positions of lighter atoms (such as H) are more difficult to ascertain. We find excellent agreement between our all bond lengths and those of the later X-ray study (Ref. 12). However, in comparing with the earlier experimental study, we find a larger difference in the C2–F bond length. The use of an ideal tetrahedral angle (109.5º) in their fitting of the X-ray data may have contributed to the differences in their atomic positions and bond lengths as compared to theory and the later experiment. Additionally, the later experiment utilized seventeen independent reflections and first layer lines to determine their structure. Only twelve reflections were used by Lando et al.[11]

## B. Spontaneous polarization

The computation of the spontaneous polarization, $P_s$, and any derived values deserves particular attention. First, as mentioned above, quantitative agreement with experimental values for the polarization in β-PVDF cannot be achieved due to low crystallinity and/or inhomogeneity of real samples as compared to the pure theoretical solid. However, differences in polarization can provide some important results as to the properties of the material at the atomic level. For example, these differences can be used to gauge whether copolymer of PVDF or boron-nitrogen substituted polymers offer enhanced properties as compared to PVDF.[30] Specifically, by determining the difference in polarization for strained and unstrained forms of the polymers, piezoelectric constants were determined and found to be substantially improved as compared to those of pure β-PVDF.

Using the theoretically determined atomic positions in Table II and the electronic wave functions obtained from density-functional calculations, the $P_s$ for the planar-zigzag structure of β-PVDF was found to be 0.181 C/m$^2$ per unit cell volume. This value is in excellent agreement with polarization found in the study by Nakhmanson et al. (0.178 C/m$^2$ per unit cell volume) for their β-PVDF structure, which used density-functional calculations within the GGA.[30] The increase in polarization by using the Berry-phase approach as compared to the rigid-dipole model (Ref. 22) indicates a large polarizing effect resulting from dipole-dipole interactions that are not present in the latter model. Since the experimentally-determined polarization values are for samples with 50% crystallinity, a very simple approximation can be made by dividing our polarization (found for a sample with 100% crystallinity) in half.[18,19,22,43] This yields a polarization of 0.091 C/m$^2$, which is remarkable close to high end of the range of values from experiment (0.05 to 0.086 C/m$^2$).

## C. Dynamic Born effective charges

As stated above, the computation of Born effective charges requires the determination of the polarization difference between a displaced and un-displaced structure. It has been shown that the $P_s$ is linear in atomic displacement to a good approximation.[44] In the case of β-PVDF, atomic displacements were made in the polarization direction of approximately 0.1% of the $b$ lattice constant. The electronic wave functions for these displaced structures were then determined using density-functional theory and the spontaneous polarization found for each structure. The spontaneous polarization was found for the un-displaced structure, using the relaxed structure from Table II. The computed $Z^*$ are contained in Table IV. The $Z^*$ satisfy the acoustic sum rule ($2 \times Z^*(C1) + 2 \times Z^*(C2) + 4 \times Z^*(F) + 4 \times Z^*(H) = 0.00$).

Several interesting observations can be made regarding the effective charges. Owing to its high electronegativity, the $Z^*$ for F (-0.76) is close to its nominal ionic valence charge (-1). It produces a highly polar bond with C2. The $Z^*$ of H (0.14) is not very close to its nominal valence charge (+1). This indicates the less polar character of the C1-H bonds as compared the C2-F bonds. The result is not surprising considering the electronegativities of C and H are similar.

The effective charges can give an indication of the dynamic charge transfer between atoms in the polymer. By examining the sum of effective charges within a given portion of the polymer, it is possible to assess the degree of electronic coupling between the dipoles in each portion. For example, by summing the effective charges within the –$CF_2$– group ($Z^*(C2) + 2 \times Z^*(F) = -0.08$) and within the –$CH_2$– group ($Z^*(C1) + 2 \times Z^*(H) = +0.08$), it shows that the C1–C2 bond has a small dipole moment and that the dipoles with the –$CF_2$– and –$CH_2$– moieties are to a large extent distinct.

To our knowledge, computed atomic charges from DFT calculations for β-PVDF have not been reported in the literature. Instead, we have included in Table IV atomic charges computed from other methods.[7] It is important to note that these charges were found from calculations on 1,1,1,3,3-pentafluorobutane, a representative molecule that mimics the β-PVDF structure. We have chosen atoms within the 1,1,1,3,3-pentafluorobutane structure that have a similar bonding environment to the corresponding atoms in the β-PVDF structure.[45]

The potential-derived charges (PDQ) were found from using the PS-Q program (Ref. 46) to determine the Hartree-Fock charge density. The charge density is then used to calculate the electrostatic potential at several thousand points outside the van der Waals radii of the molecule and then atom-centered charges are optimized to fit the electrostatic potential. Finally we include charges found from Mulliken population analysis. It is important to note that the PDQ and Mulliken methods yield static atomic charges. Therefore quantitative agreement with values determined using dynamic approaches for systems with strong covalent bonding is difficult to achieve. This lack of agreement may be caused by the inherent differences in the methods, as described above, basis-set dependency and/or the inability of the 1,1,1,3,3-pentafluorobutane to characterize the chemical bonding in β-PVDF.

There are however some comparisons that can be made from the results. By comparing the ratio of atomic charges, it is possible to assess the ability of different computational methods to describe the bonding within the system. For example,

$$\frac{Z^*(C2)}{Z^*(F)} = -1.89; \frac{PDQ(C2)}{PDQ(F)} = -2.76; \frac{Mulliken(C2)}{Mulliken(F)} = -1.96$$

$$\frac{Z^*(C1)}{Z^*(H)} = -1.43; \frac{PDQ(C1)}{PDQ(H)} = -3.01; \frac{Mulliken(C1)}{Mulliken(H)} = -2.37 \ .$$

$$\frac{Z^*(C1)}{Z^*(C2)} = -0.14; \frac{PDQ(C1)}{PDQ(C2)} = -0.75; \frac{Mulliken(C1)}{Mulliken(H)} = -0.52$$

The Mulliken charge ratios are consistently closer to the $Z^*$ ratios as compared to the ratios of PDQ charges. The ratios of static atomic charges (PDQ and Mulliken) for the more covalent bonds (C1–H and C1–C2) are considerably different that of the dynamic charges ($Z^*$) for those same bonds. This gives an indication of the inaccuracy of static atomic charge determination for covalently bonded atoms. However, the ratios of static atomic charges for the C2–F bond are much closer to the same ratio of $Z^*$. Since this bond is more ionic in character, static atomic charges will be more accurate.

## 4. CONCLUSIONS

In conclusion, we have determined relaxed atomic positions for the planar-zigzag structure for β-phase of poly(vinylidene fluoride) using experimentally-determined lattice constants. We have employed density-functional theory within the generalized gradient approximation using optimized pseudopotentials. In addition, we computed the spontaneous polarization for the structure using a Berry-phase approach. From finite differences in the polarization we computed Born effective charges for each atom in the polymer. These effective charges indicate highly polar carbon-fluorine bonds and less polar carbon-hydrogen bonds in the polymer. Finally, using the effective charges, we show that there is small amount of charge transfer between the –CH$_2$– and –CF$_2$– groups within the polymer.

ACKNOWLEDGMENTS

The authors would like to thank Andrew M. Rappe, Ilya Grinberg and Valentino Cooper for their invaluable help with the entire project.  This work was supported by a grant from the Research Committee of the C. W. Post Campus of Long Island University.  Funding for computational equipment was provided by Long Island University.

TABLES

TABLE I. Construction parameters for the C, F and H pseudopotentials. Core radii ($r_c$) are in atomic units and $q_c$ are in Ry$^{1/2}$.

| Atom | Reference configuration | $r_c$ | $q_c$ |
|---|---|---|---|
| C | $2s^2$ | 0.84 | 7.07 |
|   | $2p^2$ | 1.29 | 7.07 |
| F | $2s^2$ | 1.14 | 7.07 |
|   | $2p^5$ | 1.63 | 7.07 |
| H | $1s^1$ | 0.77 | 7.07 |

TABLE II. Theoretical and experimental atomic positions for planar-zigzag structure of β-poly(vinylidene fluoride). Positions are given as fractions of lattice constants.

| Atom | Theory[a] | | | Experiment[b] | | | Experiment[c] | | |
|---|---|---|---|---|---|---|---|---|---|
|  | x/a | y/b | z/c | x/a | y/b | z/c | x/a | y/b | z/c |
| C1 | 0.000 | 0.000 | 0.000 | 0.000 | 0.000 | 0.000 | 0.000 | 0.000 | 0.000 |
| C2 | 0.000 | 0.170 | 0.500 | 0.000 | 0.176 | 0.500 | 0.000 | 0.174 | 0.500 |
| F | 0.129 | 0.342 | 0.500 | 0.130 | 0.334 | 0.500 | 0.126 | 0.355 | 0.500 |
| H | 0.103 | -0.132 | 0.000 | n/a | n/a | n/a | 0.105 | -0.124 | 0.000 |

[a] $a = 8.58$ Å, $b = 4.91$ Å, and $c = 2.56$ Å.

[b] Reference 11 with $a = 8.47$ Å, $b = 4.90$ Å, and $c = 2.56$ Å.

[c] Reference 12 with $a = 8.58$ Å, $b = 4.91$ Å, and $c = 2.56$ Å.

TABLE III. Theoretical and experimental bond length and bond angles for planar-zigzag structure of β-poly(vinylidene fluoride) computed from reported atomic positions. Values in parentheses indicate quantities used in fitting X-ray data to determine experimental atomic positions (See Table II). Bond lengths are given in Angstroms and bond angles in degrees.

|  | Theory | Experiment[a] | Experiment[b] |
|---|---|---|---|
|  |  | Bond Lengths |  |
| C1–H | 1.10 | (1.094) | 1.09(1.09) |
| C2–F | 1.39 | 1.35(1.344) | 1.40(1.34) |
| C1–C2 | 1.53 | 1.54(1.541) | 1.54(1.54) |
|  |  | Bond Angles |  |
| H–C1–H | 107.5 | (109.5) | 111.9(112) |
| F–C2–F | 105.3 | 109.8(109.5) | 101.2(108) |
| C1–C2–C1 | 113.6 | 112.1(112.4) | 112.6(112.5) |

[a] Reference 11.

[b] Reference 12.

TABLE IV. Computed Born effective charges ($Z^*$) for the planar-zigzag structure of β-poly(vinylidene fluoride). Comparison atomic charges for 1,1,1,3,3-pentafluorobutane are also provided: PDQ (potential-derived charges) and Mulliken (charges from Mulliken population analysis). See text for description.

| Atom | $Z^*$ | PDQ[a] | Mulliken[a] |
|---|---|---|---|
| C1 | -0.20 | -0.547 | -0.406 |
| C2 | 1.44 | 0.733 | 0.781 |
| F | -0.76 | -0.266 | -0.398 |
| H | 0.14 | 0.182 | 0.171 |

[a] References 7 and 46.

FIGURES

FIG. 1. Molecular structures for β-poly(vinylidene fluoride). (a) planar-zigzag and (b) alternatively-deflected structures. See text for description.

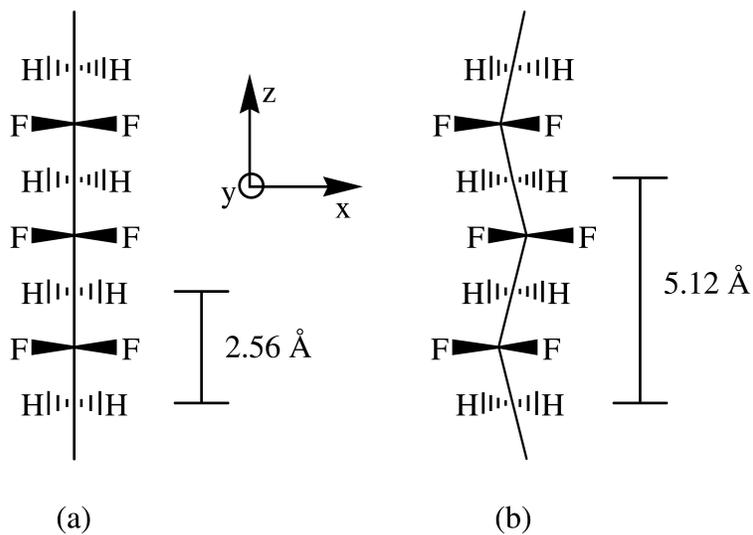

(a)  (b)